# Photoredox Processes in the Aggregation and Gelation of Electron-responsive Supramolecular Polymers Based on Viologens

Clément Roizard,[1] Vivien Andrieux,[1] Shagor Chowdhury,[1] Quentin Reynard-Feytis,[1] Christophe Kahlfuss,[1] Eric Saint-Aman,[2] Floris Chevallier,[1] Christophe Bucher,[1,z] Thomas Gibaud,[3,z] Denis Frath[1,z]

[1] Univ Lyon, Ens de Lyon, CNRS UMR 5182, Laboratoire de Chimie, F69342 Lyon, France
[2] Univ Grenoble Alpes, CNRS UMR 5250, Département de Chimie Moléculaire, F38058 Grenoble, France
[3] Univ Lyon, Ens de Lyon, CNRS UMR 5672, Laboratoire de Physique, F69342 Lyon, France
[z] Corresponding Authors E-mail Addresses: denis.frath@ens-lyon.fr, thomas.gibaud@ens-lyon.fr, christophe.bucher@ens-lyon.fr

**Abstract**

Viologen-based ditopic bis-pyridinyl-triazole bidentate ligands self-assemble in the presence of palladium ions into supramolecular polymers whose structure is imposed by the directed formation of coordination bonds. Light-irradiation of these electron-responsive supramolecular materials triggers a photo-induced electron transfer yielding isolated π-radicals and dimers of radicals. The photoreduction events and the associated dimerization steps trigger a large-scale reorganization occurring within the supramolecular network yielding aggregates or gels depending on the irradiation conditions (power, duration). Detailed electrochemical, spectro-electrochemical and photochemical analyses were conducted to understand the mechanisms at stakes in these light-induced aggregation and gelation.

**Introduction**

Soft electronics and supramolecular electronics have emerged over the past decade as cutting edge research topics with great potential for the development of flexible electronic devices.[1–7] The main obstacles to the development of these fields is the integration of soft materials or supramolecular assemblies in devices, their connection to electrodes and the controlled exploitation of their response to electrical stimulation. Such implementation requires the use of fast and reversible processes and of robust (supra)molecular architectures undergoing stable and well-defined responses to electrical potential. In line with these ambitious goals, there is currently great interest in the development of supramolecular materials whose macroscopic properties can be modulated by discrete molecular events triggered by remote stimulation. Gels are particularly well suited to meet these ambitious goals as they are made of a large amount of solvent trapped in anisotropic networks whose structure, strength and dynamic properties can be rationalized and anticipated through supramolecular engineering.[4–7] The potential of such soft materials has for instance recently been demonstrated through the development of a gel-based memristor exploiting conductivity changes associated to a sol/gel transition triggered by heating or DC bias.[7]

Metal-organic gels, also known as metallogels, are a specific class of gels incorporating metal ions as building elements. The presence of metals in the network has proved beneficial in many respects, not only to tune the strength and morphology of the gels but also to upgrade or tune their spectroscopic, rheological, biological, magnetic, catalytic, conductive and self-healing properties.[8,9] One efficient strategy that has been used to provide access to metal-organic networks endowed with gelating abilities relies on metal-driven self-assembly of tailor-made polytopic organic ligands affording coordination polymers. Thanks to the dynamic nature of the metal-ligand bonds involved in their network, such polymers are ideally suited to the development of stimuli-responsive gels capable of achieving drastic changes in shape, appearance, rheological properties or of undergoing sol/gel phase transition in response to external stimuli.[9–11] Such responsiveness to stimulation (pH, T, P, light) has for instance proved useful in catalysis or to the development of sensors, actuators, memory devices, smart material, for pollutant removal or to regulate drug delivery and adhesion.[9,12–16] The incorporation of redox/photo-active units based on N,N'-disubstituted 4,4'-bispyridinium salts, commonly known as viologens, in the structure of supramolecular gels has been shown to confer interesting electron-accepting and electrochromic properties to the materials.[17–21] These responsive units have already been used as key elements for the development of electrochromic devices, molecular machines and organic batteries.[22] Recently, viologens have also been used as redox-responsive components in molecular junctions showing great promises for uses as conductive material in molecular electronics.[23–25]

Our research efforts in this field have led us to focus on supramolecular architectures and coordination polymers involving viologens as key redox/photo-responsive building elements.[26–31] In particular, we have developed different strategies allowing to exploit the ability of viologen derivatives to form π-dimers[32] in their reduced state to achieve a remote control over their organization within supramolecular assemblies.[26–31,33,34] Among our recent achievements involving metallogels, we have reported a sol/gel transition triggered by light-irradiation of a viologen-based coordination polymer formed in the presence of palladium ions.[30] Following these initial findings, we have carried out detailed microscopy and rheological investigations to provide further insights into the light-triggered gelation mechanism.[31] On the ground of time-resolved and micro-structure analyses, we were able to propose a unique physical pathway driven by an arrested phase separation leading to aggregation or gelation. At low irradiation time and lamp power, the sample was found to form aggregates. The aggregated sample is spatially heterogeneous and composed of individual elongated aggregates randomly oriented of a few tens of microns in size. All the investigated gels display the same structural and mechanical properties, the main characteristic being a spatially homogeneous and weakly contrasted network with a correlation length $\xi \simeq 7$ μm and a weak elastic modulus $G' \simeq 0.3$ Pa.

However, many aspects of this complex light-triggered phase transition remained to be studied: the photo-induced processes leading to the formation of viologen cation radicals, the structure of the coordination polymer in its reduced states (box, folded chain, inter-chain π-dimers), the interactions between building blocks at the reduced state (orbital overlap of viologen radical cations, metal–metal bonds) and the correlation between changes occurring at the molecular scale (chemical pathway) and those observed at the micro/macroscopic scale (physical pathway). We now report in this article further chemical and physical data allowing to improve the understanding of these phenomena.

**Experimental**

All solvents and reagents were obtained commercially unless otherwise noted. Dimethylformamide (DMF, Sigma Aldrich, extra-dry with molecular sieves, water < 0.01%) was used for all the studies. The tetra-*n*-butylammonium perchlorate salt used for (spectro)electrochemical studies was purchased and used without further purification (TBAP, Fluka puriss.). [Pd(CH$_3$CN)$_4$](BF$_4$)$_2$ was obtained from Acros Organics.

Compound **1**(PF$_6$)$_4$ and **7**(PF$_6$)$_2$ were prepared following procedures described previously.[30] Supramolecular polymer **2**$^{n+}$ was obtained *in-situ* by mixing equimolar amounts of **1**$^{4+}$ and [Pd(CH$_3$CN)$_4$](BF$_4$)$_2$ in DMF. Coordination complex **8**$^{4+}$ was formed *in-situ* by mixing two equivalents of **7**$^{2+}$ and one equivalent of [Pd(CH$_3$CN)$_4$](BF$_4$)$_2$ in DMF. Acquisition and analysis of the cyclic voltammetry (CV) and spectro-electrochemical (SEC) data for those (supra)molecular structures were reported before.[30]

Photoreduction experiments were performed using Thorlabs M365L2 ($\lambda_{max}$ = 365 nm), or M455L3 ($\lambda_{max}$ = 455 nm) mounted high-power LED lamp collimated with a Thorlabs LA1951 Plano-convex lens (f = 25.4 nm). White-light irradiations were conducted using an unfiltered X-Cite 120 LED boost from Excelitas (spectrum shown in Figure S1) placed directly above the sample or through the 10x objective in the reflection mode of the microscope (Nikon Plan fluor, N.A.=0.3). All samples were prepared under an argon atmosphere (glove box) and introduced into a dismantled quartz absorption cuvette featuring a 0.1 mm layer thickness. Irradiation was conducted outside the glovebox on samples placed at the lamp focal distance. Absorption spectra were collected with a MCS 500 or MCS 601 UV-NIR Zeiss spectrophotometer.

Microscopy experiments were carried out on a Nikon Ti-eclipse inverted light microscope in bright field. We used a 20× objective to image the sample. The images of the sample were recorded with a camera (ORCA-Flash4 Digital CMOS camera from Hamamatsu) at time t~2000 s after irradiation when the structure no longer evolves.

**Results**

The ditopic ligand **1**(PF$_6$)$_4$ used as the building block of the coordination polymer is depicted in Scheme 1. The latter features two planar binding units (triazole/pyridine) introduced on each side of a viologen-based mechanical hinge that can reversibly alternate under stimulation between a folded and an elongated conformation. This ligand was found to self-assemble in the presence of palladium ions ($Pd^{2+}$) to yield linear coordination polymers.[30] This ligand/palladium pair was selected for the following reasons: i) palladium ions are well stabilized by nitrogen-based ligands capable of providing $PdN_4$ square-planar environments, ii) the nitrogen–palladium bond is known to be quite labile, and iii) palladium complexes are usually electrochemically inert over a quite large potential window.[35] Due to its chelating $N_2$ structure, the terminal triazole/pyridine ligand was moreover selected for its ability to form a well-defined/stable square-planar Pd(II) complex.[36] The formation of coordination polymers from equimolar mixtures of **1**$^{4+}$ and $Pd^{2+}$ was initially established from DOSY NMR measurements. Further investigations revealed that the resulting coordination polymer **2**$^{n+}$ (Scheme 1) is capable of undergoing large-scale reorganizations in solution under light or electrical stimulation,[30] the driving force being the non-covalent association (π-dimerization) of the *in-situ* generated viologen radical-cations ($V^{+\bullet}$).[26–34] The exact nature of the π-dimerized structure(s) has so far remained quite hypothetical, but all the data collected suggest the formation of box-shaped discreet assemblies (**4** in Scheme 1), of folded polymers (**5** in Scheme 1) or of inter-chain dimers[37] (**6** in Scheme 1) in the gel phase. In order to decipher the photoredox processes involved in the light induced aggregation and gelation mechanisms, detailed electrochemical, spectro-electrochemical and photochemical investigations were conducted both on the isolated ligands and on the viologen-based coordination polymer.

*Electrochemical studies.* The cyclic voltammetry (CV) curves of the free ligand **7**$^{2+}$ shows two reversible reduction waves at $E_{1/2}^1 = -0.670$ V ($\Delta E_p^1 = 66$ mV at $v = 0.1$ V s$^{-1}$) and $E_{1/2}^2 = -0.980$ V ($\Delta E_p^2 = 64$ mV at $v = 0.1$ V.s$^{-1}$) (Figure 1). These waves were attributed to the successive formation of the viologen-based cation radical **7**$^{+\bullet}$ and neutral quinonic species **7**$^0$ (solid line in Figure 1B). Similar reversible waves centered on viologens have been observed for **1**(PF$_6$)$_4$, but within a quite different potential range (Figure 1A). The first reduction, leading to the bis(radical-cation) **1**$^{2(+\bullet)}$, is observed at $E_{1/2}^1 = -0.566$ V ($\Delta E_p^1 = 42$ mV at $v = 0.1$ V s$^{-1}$) while the second ill-defined reduction wave corresponding to the formation of the quinonic species **1**$^0$ appears at $E_{pc}^2 = -1.054$ V. The unusual shape of the second reduction wave is attributed to the adsorption of **1**$^0$ onto the electrode surface, a process which is revealed by the intense desorption peak centered at $E_{pa} = -0.538$ V observed on the reverse scan (this adsorption process can be avoided and the reversibility of this reduction wave can be recovered upon carrying out these analyses in less concentrated solutions $\leq 0.2$ mM). Comparing the signature of both compounds leads to the conclusion that **1**$^{4+}$ is more easily reduced than **7**$^{2+}$ while **1**$^{2(+\bullet)}$ is harder to reduce than **7**$^{+\bullet}$. As a matter of fact, the large stability domain of the bis-radical species **1**$^{2(+\bullet)}$ ($\Delta E_{pc} \sim 480$ mV to be compared to $\Delta E_{pc} \sim 310$ mV measured for the reference compound **7**$^{+\bullet}$), together with the remarkably low $\Delta E_p^1$ value measured on its first reduction wave (42 mV) are both resulting from the existence of a π-dimerization process coupled to the electron transfer yielding the intramolecular π-dimer $[\mathbf{1}^{2+}]_{Dim}$.

Binding of palladium with ligand $\mathbf{1}^{4+}$ and $\mathbf{7}^{2+}$ has also been analyzed by electrochemical methods (Figure 1C and 1D). We found that the CV curves recorded for the reference compound $\mathbf{7}^{2+}$ in the presence of 0.5 molar equivalent of $[Pd(CH_3CN)_4](BF_4)_2$ (dotted line in Figure 1D; $E_{1/2}^1$ = -0.667 V, $\Delta E_p^1$ = 56 mV) are almost identical to those measured with $\mathbf{7}^{2+}$ alone ($E_{1/2}^1$ = –0.670 V, $\Delta E_p^1$ = 66 mV) (solid line in Figure 1D). This good matching suggests that coordination of palladium to the ligand has a negligible effect on the electronic properties of the viologens and that the cation radical species $\mathbf{7}^{+\bullet}$ are not involved in π-dimerization processes in these experimental conditions. Addition of 1 molar equivalent of $Pd^{2+}$ to a solution of $\mathbf{1}^{4+}$ (1 mM in DMF + TBAP 0.1 M) led to drastic changes in the shape and potential of the first reduction wave (Figure 1C). It includes a large drop of the peak intensity coming along with a slight anodic shift of the peak potential ($\Delta E$ ~ 20mV) associated to a loss of reversibility. As a general statement, the peak intensity measured for a given CV wave depends on the concentration and on the square root of the diffusion coefficient of the electroactive species.[38] Such direct link between current and diffusion can be raised to account for the significant loss of intensity observed after addition of palladium, since the diffusion coefficient is expected to decrease upon formation of metal-organic self-assembled species in solution. Other arguments had to be put forward to explain the 20 mV shift of the peak potential value towards less negative values and the irreversible shape ($\Delta E_p$ =135 mV at $v$ = 0.1 V s$^{-1}$) of the wave recorded in the presence of $Pd^{2+}$ (full line in Figure 1C). The fact that reduction occurs at a slightly less negative potential than the free ligand reveals that the electron-triggered dimerization proceeds as efficiently in both cases, the 20 mV anodic shift being only due to the electron-withdrawing effect of the palladium center on the appended viologens (a similar shift was observed on Figure 1D for the reference compound). The most striking finding is that the re-oxidation peak observed on the reverse scan undergoes a large anodic shift (~+135 mV at $v$ = 0.1 V s$^{-1}$) upon addition of metal. We found of course that the amplitude of the $\Delta E_p$ value evolves with scan rate or with temperature. For instance, $\Delta E_p$ happens to rise from 125 to 186 mV when scan rate is increased from 50 to 750 mV s$^{-1}$ and heating the electrolytic solution triggers a decrease of the $\Delta E_p$ value going from 135 mV (1 mM, $v$ = 0.1 V s$^{-1}$, 24 °C) to 93 mV (70 °C). In all cases, the re-oxidation of the π-dimerized metal-organic structure(s) generated in solution occurs at much less negative potential when palladium is present in solution, which means that the π-dimer complexes are much more stabilized in the presence than in the absence of metal. All the experimental evidences discussed above thus support the conclusion that π-dimerization of the viologen radicals, coming along with a folding of the tweezer-like tectons, occurs efficiently in the self-assembled coordination polymer and that complexation of palladium further stabilizes the dimerized structures.

***Spectro-Electrochemical Characterizations.*** The ability of $\mathbf{1}^{4+}$ or $\mathbf{7}^{2+}$ to form non-covalent π-dimers in organic solution has been further investigated by spectro-electrochemical analyses, which involved regularly recording absorption spectra over time during potentiostatic reductions carried out in DMF electrolyte at platinum electrodes (Figure 2). The folding motion yielding the intramolecular dimer $[\mathbf{1}^{2+}]_{Dim}$ is revealed on the UV/Vis spectra recorded during the exhaustive reduction (1 electron per viologen subunit, $E_{app} = -0.8$ V) of $\mathbf{1}^{4+}$ through the growth of diagnostic absorption bands at 403 nm (74200 L.mol$^{-1}$.cm$^{-1}$), 565 nm (44000 L.mol$^{-1}$.cm$^{-1}$) and 910 nm (12700 L.mol$^{-1}$.cm$^{-1}$) (Figure 2A). This signature contrasts with that recorded in the same experimental conditions with the reference compound $\mathbf{7}^{2+}$, whose reduction leads to a different set of absorption bands attributed to a simple non associated viologen based cation radical (422 nm (26300 L mol$^{-1}$ cm$^{-1}$), 714 nm (5400 L.mol$^{-1}$.cm$^{-1}$), and 640 nm (9700 L.mol$^{-1}$.cm$^{-1}$) (Figure 2C). The differences observed between Figure 2A and 2C thus confirms the intramolecular character of the π-dimerization leading to $[\mathbf{1}^{2+}]_{Dim}$.

The UV/Vis spectra recorded during the exhaustive reduction (1 electron per viologen, $E_{app}$ = -0.8 V) of the reference 1/2 (M/L) palladium complex $\mathbf{8}^{4+}$, generated in situ from $\mathbf{7}^{2+}$ and 0.5 molar equivalent of [Pd(CH$_3$CN)$_4$](BF$_4$), shows here again the characteristic fingerprint of a viologen-based radical cations (Figure 2D). The absorption bands growing at 421 nm (20600 L.mol$^{-1}$.cm$^{-1}$) and 714 nm (5400 L.mol$^{-1}$.cm$^{-1}$), as well as the broad signal centered at 640 nm (9700 L.mol$^{-1}$.cm$^{-1}$) are thus attributed to the formation of the reduced complex $\mathbf{8}^{2(+\bullet)}$. The great similarity observed between the curves shown in Figure 2C and Figure 2D, taken together with the absence of band in the near IR region, led us to conclude that the Pd$^{2+}$ ion precludes the dimerization of both viologen cation radicals.

The spectra recorded over the course of the exhaustive reduction ($E_{app}$ = -0.8 V) of the self-assembled polymer $\mathbf{2}^{n+}$, produced *in-situ* from an equimolar mixture of $\mathbf{1}^{4+}$ and [Pd(CH$_3$CN)$_4$](BF$_4$)$_2$, are shown in Figure 2B. The first stage of the electrolysis, up to ~1 electron exchanged per molecule, involves the progressive growth of absorption bands at 400 nm and 586 nm coming along with a broad signal centered at 930 nm (solid line in Figure 2B). Further reduction of the solution, up to 2 electrons exchanged per molecule, then led to the progressive drop in the intensity of these initial signals at the expense of new bands growing at 393 nm and 568 nm, while the NIR absorption band gained both in intensity and broadness (dashed lines in Figure 2B). This complex behavior reveals the existence of multiple equilibria involving dimerized and non-dimerized viologen radicals. The UV-Vis absorption spectra recorded in the early stage of the bulk electrolysis clearly reveal the simultaneous presence of free viologen radicals ($\lambda_{max}$ = 400 nm and 586 nm) and of π-dimerized radicals ($\lambda_{max}$= 900 nm and shoulder at ~ 570 nm). This situation then evolves upon completing the electrolysis (from 1 to 2 electron/molecule) towards full consumption of the free radical in favor of π-dimerized structures ($\lambda_{max}$ = 393, 568 and 900 nm). The broadness of the near IR band is moreover consistent at this stage with the existence of different types of dimers in the polymers. The mechanism shown in Scheme 1 relies on the assumption that the viologen-based hinges located at both ends of each self-assembled oligomer will have higher degrees of freedom, and thus higher dimerization kinetics, compared to those embedded in the structure. The first molecule to dimerize are thus those located on these terminal positions to form partially folded structures **3** (Scheme 1). Then on longer time scales, when all the viologen units get finally reduced, all the equilibria involving radicals are progressively displaced towards a mixture of macrocyclic structures **4** or completely folded polymeric assemblies **5** (Scheme 1), as revealed by the unusual broadness of the near IR absorption band recorded by the end of the experiment.

*Photochemical studies*. The photoinduced sol→gel transition was initially induced using a white-light source (the lamp spectrum is provided in the ESI section, Figure S1) in presence of Ru(bpy)$_3$ and triethanolamine (TEOA) used as photosensitizer and sacrificial donor respectively.[30] The exact photochemical mechanism remained however unclear. We therefore conducted additional investigations to better understand the chemical pathway triggered by photoirradiation. We first performed a series of photo-reduction experiments which involved submitting polymer **2**$^{n+}$ to irradiation using monochromatic LED lamps centered at 365 nm or 455 nm (Figure 3). These preliminary studies revealed that an intramolecular photo-induced electron transfer[39] (PET) occurs when irradiating the samples in the UV bands of the triazole/pyridine ligands at 365 nm (Scheme 2). The studies revealed also that in our conditions, when irradiating at 455 nm, the Ru(bpy)$_3$ photosensitizer does not significantly contribute to the photoreduction process.

As can be seen in (Figure 3A), irradiation of self-assembled polymer **2**$^{n+}$ at 365 nm led to a set of changes with striking similarities to those observed during the spectro-electrochemistry experiment and characteristic of π-dimers formation (Figure 2B). A photo-stationary state was obtained after approximately 10 minutes of irradiation, which correspond to the photoreduction of about 36% of the viologen subunits involved in the polymer (blue curve in Figure 4B). Similar experiments conducted on the reference compound 1,1-dimethyl-4,4'-bipyridinium revealed that photoreduction does not proceed efficiently upon irradiation of the viologen unit (π-π*) at 365 nm. As can be seen in Figure S2, the absorbance attributed to the photogenerated radical cation increases progressively until reaching a maximum intensity after 30 minutes of irradiation corresponding to a conversion of only 6%. We also found that this signal sees its intensity decrease during prolonged irradiation due to photobleaching. All these measurements thus clearly demonstrate that the PET mainly relies on the formation of the excited state of the conjugated triazole/bipy unit and not on the formation of excited viologens

We also discovered that similar changes, proceeding more slowly, can be observed when irradiating samples of **2**$^{n+}$ at 455 nm in the absence (green curve in Figure 3B and Figure S3) or in the presence of Ru(bpy)$_3$ (red curve in Figure 3B and Figure S4). The reduction for irradiation at 455 nm was found to be slightly faster in the absence of ruthenium complex, probably due to the absence of competitive photochemical pathways. The poor efficiency of the photosensitizer Ru(bpy)$_3$ can in fact be explained by the bi-molecular nature of the intermolecular photoactivation process with respect to the intramolecular character of PET activated by irradiation of the triazole/pyridine ligands. Quenching of the Ru(bpy)$_3$ photosensitizer excited state due to the presence of the palladium center could also play a role.

A faster photoreduction kinetics was indeed observed when irradiating at 455 nm a solution of ligand **1**$^{4+}$ alone in the absence of palladium ion and Ru(bpy)$_3$ (orange curve in Figure 4B and Figure S5), with 11% of conversion after 20 minutes versus 6% obtained with **2**$^{n+}$ in the same conditions (red curve in Figure 3B and Figure S4). The fact that PET occurs upon irradiation at 455 nm in absence of Ru(bpy)$_3$ can be explained by the small spectral overlap between the excitation source and the feet of the band attributed to the triazole/pyridine ligand from 420 nm to 500 nm. This assumption was further confirmed by showing that use of a 450-490 nm band-stop filter leads to lower efficiency (purple curve in Figure 3B and Figure S6).

We have also conducted irradiation experiments on the reference ligand **7**$^{2+}$ and on the corresponding coordination complex **8**$^{4+}$. The Photo-reduction of **8**$^{4+}$ (20 mM) at 365 nm led as expected to the bis-radical **8**$^{2(+\bullet)}$ as the main product and also to little amount of the intermolecular assembly [**8**$^{2+}$]$_{dim}$ involving two π-dimers (Figure 3C). The importance of this double dimerization on either sides of the metal center is highlighted by the fact that almost no π-dimer is observed upon irradiating the metal-free ligand **7**$^{2+}$ in the same experimental conditions (Figure 3D). These experiments also suggest that Pd-Pd interactions could be involved in the stabilization of complex [**8**$^{2+}$]$_{dim}$, and by extension in the stabilization of the supramolecular assemblies **4**, **5** and **6**.[40,41] It should also be mentioned that photobleaching was observed in both experiments after a few minutes of irradiation (dotted lines in Figure 3C and 3D).

We then focused on the photo-redox processes occurring during aggregation and gelation. Unfortunately, the gel state could not be obtained by irradiation of the samples with the 365 nm LED lamp, probably because the light intensity was not high enough. The importance of the light power on the observed macroscopic changes was already discussed in a previous report: using a white-light LED (Figure S1), aggregates were obtained at low power or for short irradiations times while sol→gel transitions were triggered by high-power light and longer irradiation times.[31]

The kinetic evolution of the spectral properties observed during the formation of aggregates is depicted in Figure 4A. The studies involved submitting samples of polymer **2**$^{n+}$ to a white-light irradiation for two minutes using a LED lamp placed directly above the sample and monitoring for approximately 30 minutes the resulting photoreduction of the viologen units by UV-Vis absorption spectroscopy (see experimental section and Figure S1). Irradiation readily led to the development of a broad band centered at 950 nm attributed to the formation of intramolecular π-dimers in the partially reduced polymer chains (structure **3** in Scheme 1). On longer time scales (usually after 2 or 3 minutes), the system spontaneously evolved with the appearance of a new band growing at ~1250 nm. The red shift can be attributed to the progressive reorganization of the reduced viologens into supramolecular structures like **4** or **5**, stabilized by multiple intermolecular π-dimerizations or by delocalization of the radical over a larger number of viologen units. This behavior was already suggested to occur in other types of supramolecular systems involving π-dimerization of viologens.[42,43]

High light intensity could be focalized on the samples upon conducting the irradiation through the objective of the microscope. Submitting samples of polymer $2^{n+}$ to such intense irradiation led to a gelation of the samples, as proved by microrheological measurements carried out on these samples (Figure 4D).[31] Unfortunately recording the time-dependent evolutions of the spectroscopic data could not be achieved in these conditions. We found that the UV-vis spectrum of the final gel state obtained after 2 minutes of irradiation is characterized by a broad band in the near infrared region (Figure 4B). As discussed above, this band can be attributed to a mixture of supramolecular assemblies (Scheme 1) but the large correlation length observed for the gel by microscopy vouches for the preferential formation of the larger-sized assemblies **6** stabilized by inter-chain interactions as discussed hereafter.

**Discussion**

In a previous report, we have shown that liquid samples of the viologen-based supramolecular polymer $2^{n+}$ in DMF can be converted into aggregates or gels depending on the photo-activation conditions.[31] The physical pathway that was proposed involves an interplay between phase separation and dynamical arrest (Scheme 3). We hypothesized that the irradiation conditions determine the type of interactions involved between the building blocks and thus their organization within the resulting materials. Here we discuss the correlation between changes occurring at the molecular and supramolecular scale upon exposure to light-irradiation (chemical pathway) and those observed at the micro/macroscopic scale (physical pathway).

*Aggregation mechanism.* According to the proposed physical pathway, aggregation is expected to occur for low irradiation power or time, namely, when the number of reduced species is still limited and the attraction between building blocks is weak. As shown in Scheme 3, it corresponds to a region of the state diagram located between the binodal line and the spinodal line. In this area, phase separation proceeds through the formation of droplets composed of a dense fluid dispersed in a dilute fluid phase. Usually, the phase separation leads to the coexistence of two homogeneous phase fluids separated by an interface. Here, aggregates result from an arrested phase separation. As the concentration inside the droplets increases and reaches the glass line, the phase separation is stopped and the droplets become isolated aggregates as observed experimentally.

Our hypothesis for the corresponding molecular scale organization is that the dense droplets forming the aggregates are composed of the rod-shaped supramolecular structures **4** favoring a nematic liquid-crystal phase (LC on Figure 4C). This is supported by the spindle-like structures observed for the aggregates. This structure is characteristic of LC dispersion observed for isotropic–nematic transition.[44,45] This type of isotrope-LC transition is a phase separation and therefore is compatible with the arrested phase separation scenario. The spindle-like structures are called tactoids and correspond to nematic droplets composed of elongated molecules that tend to point in the same direction but do not have any positional order. The spindle-like shape results from the molecule orientation and the interplay between the interfacial tension and the splay and bend elastic constants of the nematic domain.[46]

Viologen-based LC phases have recently been observed but tactoids ones have never been reported so far maybe due to their transient nature.[47–49] In our system, the preferential orientation between the building blocks leading to tactoids could be explained by the dynamic intermolecular interactions between neighboring box-shaped structures **4** based on orbital overlap of the radical over several viologen units (Figure 4C). This hypothesis is consistent with the wavelength shift observed in Figure 4A and is also supported by the fact that such intermolecular interactions where observed in the formation of the supramolecular bis-π-dimer complex [$8^{2+}$]$_{dim}$ (Figure 3C).

***Gelation mechanism.*** According to the proposed physical pathway, upon irradiation of supramolecular polymer $2^{n+}$ (Figure 4D) gelation is expected to occur for high irradiation power or time, namely, when the number of reduced species is large and the attraction between building blocks is strong. As shown in Scheme 3, it corresponds to a region of the state diagram located below the spinodal line.[50,51] At sufficiently high attraction and concentration in π-dimers, the dispersion undergoes spinodal decomposition, leading to a bicontinuous network composed of a dilute and a concentrated fluid phase. Usually, the spinodal decomposition proceeds and leads to the coexistence of two homogeneous fluid phases separated by an interface. Here, however, the glass transition interferes with the spinodal decomposition and leads to gelation. In this scenario, the concentrated network phase becomes denser meets the glass line, becomes kinetically arrested and freezes the phase separation process, leading to a gel with a large correlation length ξ as observed experimentally.

In our system, sol→gel transition occurs for high-power light and longer irradiation times. Our hypothesis to correlate phenomenon at the molecular scale with the physical pathway described above is that irradiation leads to the formation of inter-chain interactions **6** and thus to the creation of a denser network of polymer chains in which reorganization of the materials becomes difficult (Figure 4D). If irradiation is not intense/long enough, the local concentration in viologen radicals will not enable the formation of supramolecular structures like **6** and the system will have time to reorganize into species such as **4** thus favoring the formation of aggregates instead of gels. Such difference can explain why we failed to obtain a kinetic evolution of spectroscopic properties for aggregates but not for gels (Figure 4). The proposed aggregation mechanism involving intermolecular radical/radical interactions between neighboring structures **4** could also explains why the NIR band observed for the aggregates is significantly broader and red shifted than that of the gels where radicals are only delocalized between two viologens in inter-chain interactions **6**.

**Conclusions**

On the ground of detailed electrochemical, spectro-electrochemical and photochemical analyzes, we have proposed a chemical pathway explaining the aggregation or gelation processes observed upon irradiation of a palladium-containing coordination polymer. This chemical pathway could also be correlated with the physical pathway described earlier. This pathway is driven by the reduction of viologen-based building block triggered by photo-induced electron transfer from the excited state of the triazole/pyridine ligands. Our studies provide key insights into the coupled photo and chemical mechanisms at stake.

To improve and facilitate our understanding of the gelation process, rather than carrying out multiple experiments due to the large variety of techniques needed to test the gelation mechanism, we aim to carry out in a single experiment an *in-situ* analyses allowing us to simultaneously probe the chemical and physical gelation pathway. This approach is especially relevant in conductive and electron-responsive viologen-based materials for application in molecular electronics where the gelation mechanisms at play are complex. To address this issue we are currently developing a microscopic platform allowing to image the structure of the materials at the microscale and to carry out simultaneous micro-rheological, electrochemical, and UV-Vis-NIR spectroscopic measurements using home-made dedicated cell designed to stimulate samples using temperature, light and electricity.

**Acknowledgments**

The authors thank the Ecole Normale Supérieure de Lyon (ENSL) and the Centre National de la Recherche Scientifique (CNRS) for financial, logistical, and administrative supports. C.R. thanks ILM and IDEXLYON for the COLUMN internship grant. V. A. thanks ENSL for the CDSN PhD Grant. This work was supported by the LABEX iMUST (ANR-10-LABX-0064) of Université de Lyon within the program "Investissements d'Avenir" (ANR-11-IDEX-0007) operated by the French National Research Agency (ANR), by the ANR (ANR-CE06-0020-01) and by the "Mission pour les initiatives transverses et interdisciplinaires" MITI CNRS.

**Figures**

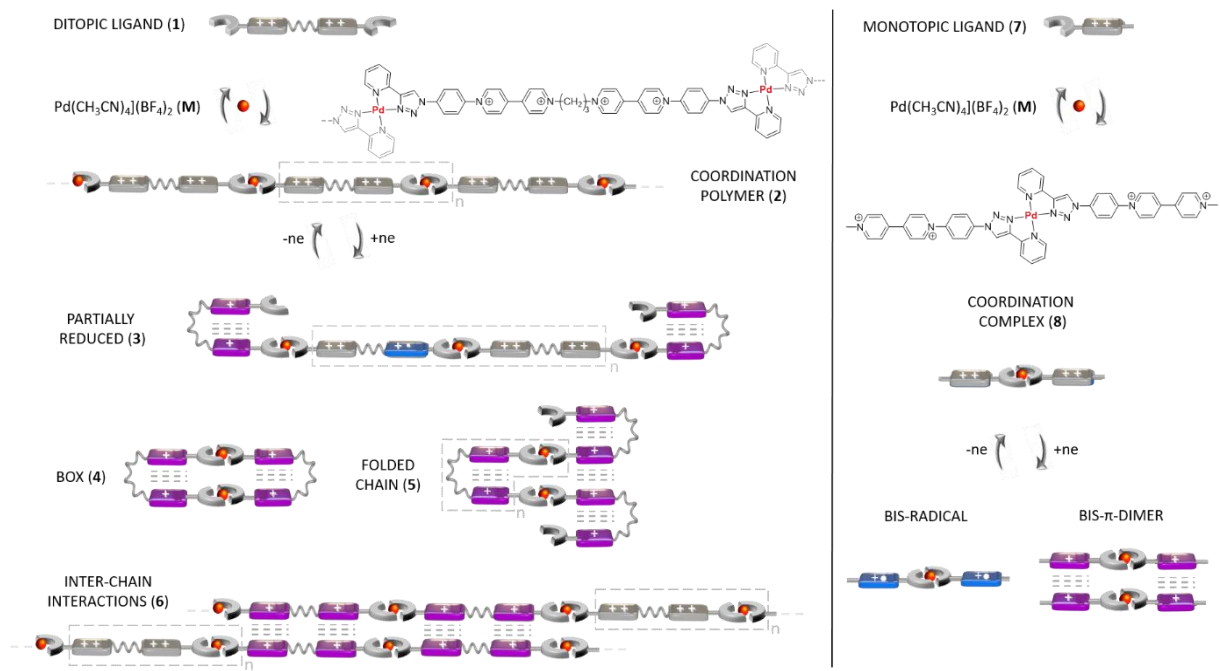

**Scheme 1.** Schematic representation of the electron-responsive materials used in this study.

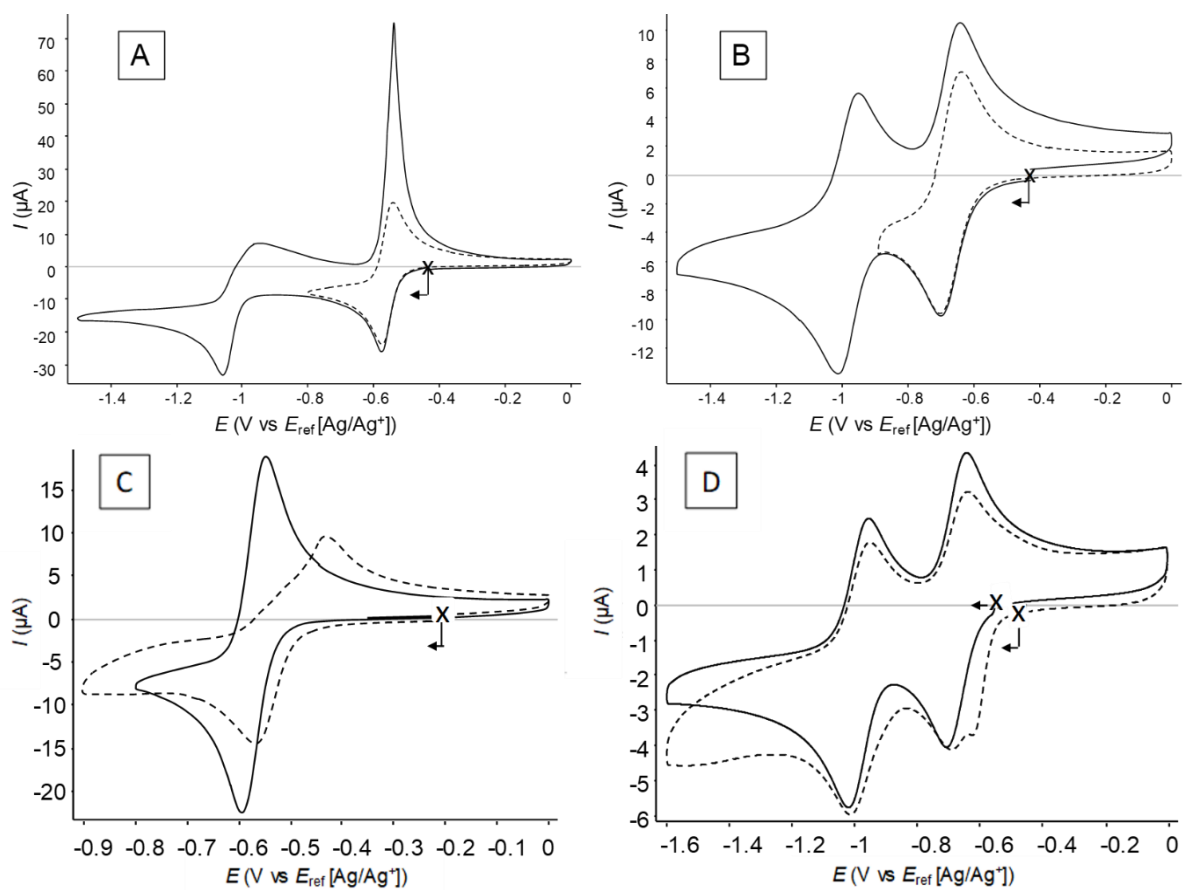

**Figure 1.** Voltametric curves recorded for A) **1** and B) **7** in DMF+ TBAP 0.1 M (1 mM, VC, Ø = 3 mm, $E$ vs Ag/Ag$^+$ 10$^{-2}$ M, $v$ = 0.1 V.s$^{-1}$). C) Voltametric curves recorded for **1** (solid line) and **2** (dotted line), (1 mM, $v$ = 0.1 V.s$^{-1}$). D) Voltametric curves recorded for **7** (solid line) and **8** (dotted line) (0.4 mM, $v$ = 0.1 V.s$^{-1}$) (DMF + TBAP 0.1 M, VC Ø = 3mm, $E$ vs. Ag/Ag$^+$ 10$^{-2}$ M). The weak reduction wave around -0.62 V is attributed to the reduction of free [Pd(CH$_3$CN)$_4$](BF$_4$)$_2$.

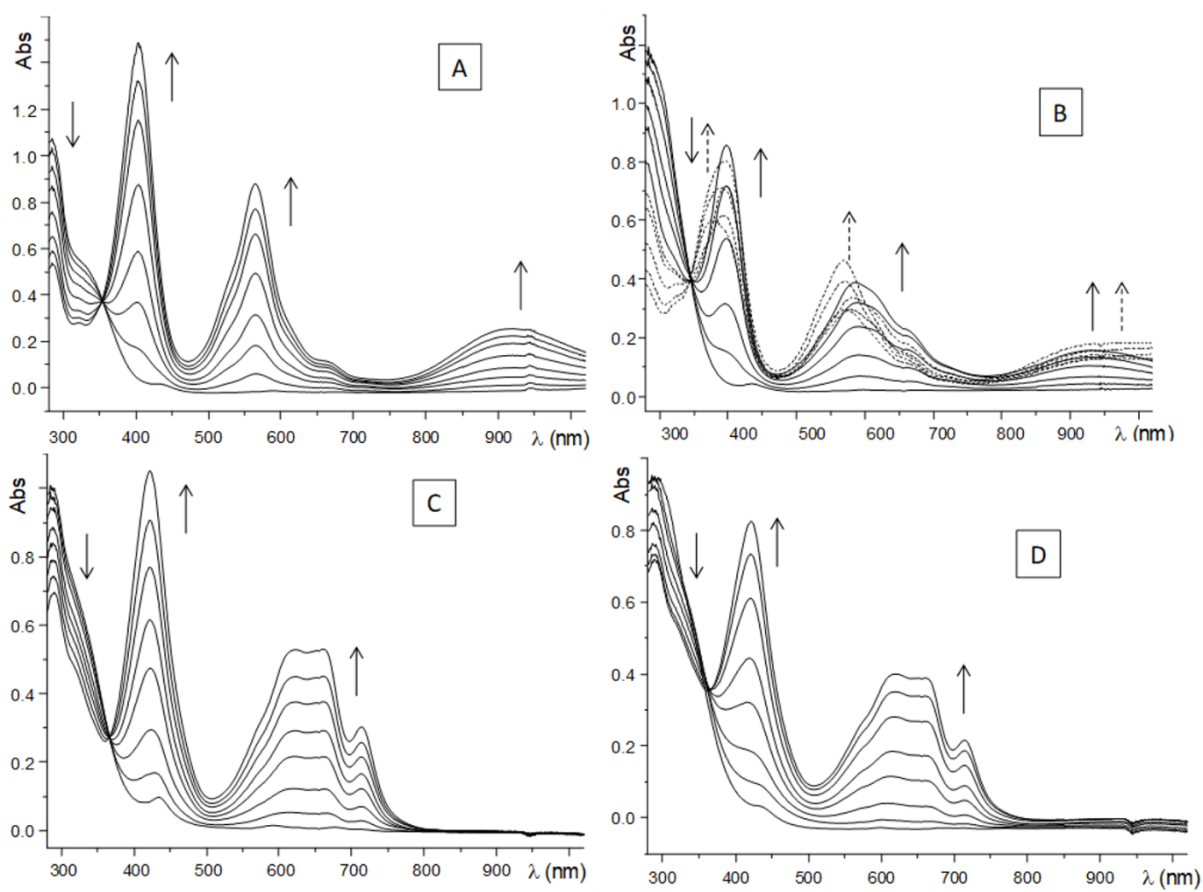

**Figure 2.** Superposition of UV/Vis spectra recorded during the exhaustive reduction (one electron per viologen subunit) of A) **1**, B) **2**, C) **7** and D) **8** (4 10$^{-4}$ mol L$^{-1}$ in viologen subunits, DMF + TBAP 0.1 M, $E_{app}$ = -0.8 V, $l$ = 1 mm).

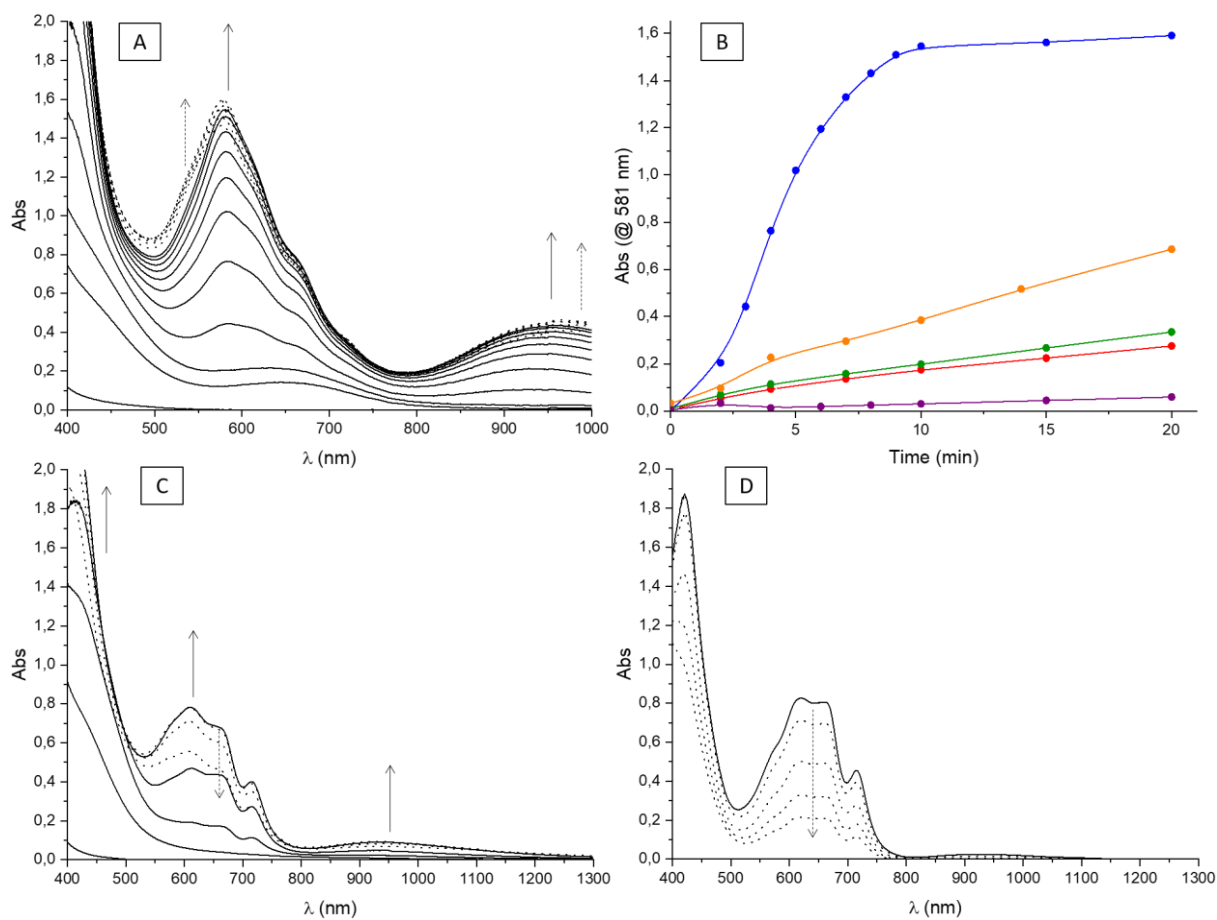

**Figure 3.** A) Superposition of UV/Vis spectra recorded over 60 minutes during the photo-reduction (365 nm) of **2** (4 $10^{-2}$ mol.L$^{-1}$ in viologen subunits, DMF + TEOA 1 $10^{-2}$ mol.L$^{-1}$, $l$ = 0,1 mm), B) Evolution of the absorbance recorded at λ = 581 nm: (blue) photo-reduction (365nm) of **2** (4 $10^{-2}$ mol.L$^{-1}$ in viologen subunits, DMF + TEOA 1 $10^{-2}$ mol.L$^{-1}$, $l$ = 0,1 mm) in absence of Ru(bpy)$_3$, (orange) photo-reduction (455nm) of **1** (4 $10^{-2}$ mol.L$^{-1}$ in viologen subunits, DMF + TEOA 1 $10^{-2}$ mol.L$^{-1}$, $l$ = 0,1 mm) in presence of Ru(bpy)$_3$, 5.5 $10^{-6}$ mol.L$^{-1}$ (green) photo-reduction (455nm) of **2** (4 $10^{-2}$ mol.L$^{-1}$ in viologen subunits, DMF + TEOA 1 $10^{-2}$ mol.L$^{-1}$, $l$ = 0,1 mm) in absence of Ru(bpy)$_3$, (red) photo-reduction (455nm) of **2** (4 $10^{-2}$ mol.L$^{-1}$ in viologen subunits, DMF + TEOA 1 $10^{-2}$ mol.L$^{-1}$, $l$ = 0,1 mm) in presence of Ru(bpy)$_3$, 5.5 $10^{-6}$ mol.L$^{-1}$ (purple) photo-reduction (455nm + 450-490 nm bandpass filter) of **2** (4 $10^{-2}$ mol.L$^{-1}$ in viologen subunits, DMF + TEOA 1 $10^{-2}$ mol.L$^{-1}$, $l$ = 0,1 mm) in absence of Ru(bpy)$_3$, C) Superposition of UV/Vis spectra during the photo-reduction (365 nm) of **8** (4 $10^{-2}$ mol.L$^{-1}$ in viologen subunits, DMF + TEOA 1 $10^{-2}$ mol.L$^{-1}$, $l$ = 0,1 mm) D) UV/Vis spectra after 2 min irradiation (solid line) and progressive photo-bleaching (dotted line) during the photo-reduction (365 nm) of **7** (4 $10^{-2}$ mol.L$^{-1}$ in viologen subunits, DMF + TEOA 1 $10^{-2}$ mol.L$^{-1}$, $l$ = 0,1 mm).

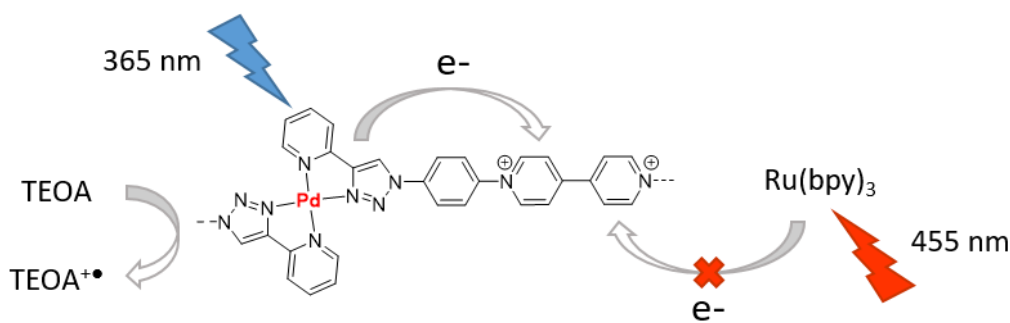

**Scheme 2.** Schematic representation of the viologen reduction photochemical mechanism.

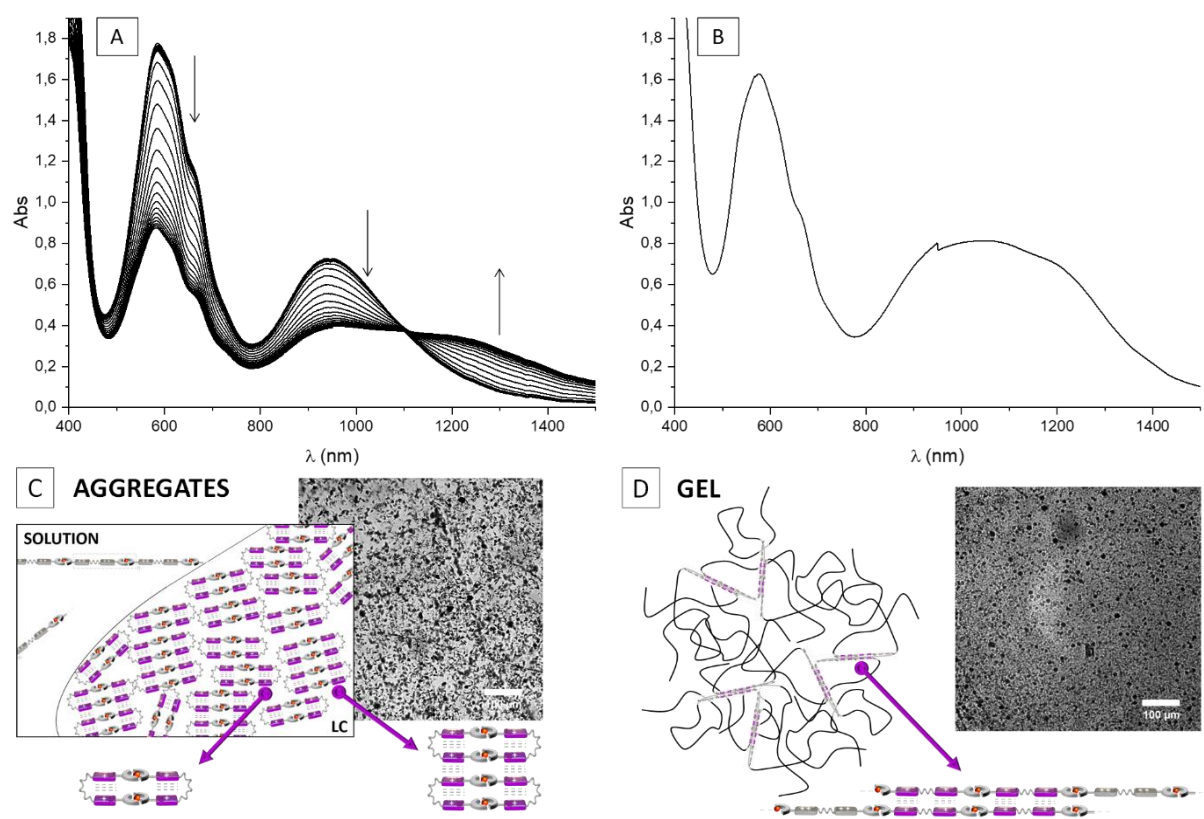

**Figure 4.** A) Superposition of UV/Vis spectra recorded during the spontaneous evolution after an initial two-minutes irradiation (direct irradiation with the white-light LED) of **2** (4 $10^{-2}$ mol.L$^{-1}$ in viologen subunits, DMF + TEOA 1 $10^{-2}$ mol.L$^{-1}$, $l$ = 0,1 mm), B) UV/Vis spectra recorded after an initial two-minutes irradiation (white-light LED through the 10x objective of the microscope) of **2** (4 $10^{-2}$ mol.L$^{-1}$ in viologen subunits, DMF + TEOA 1 $10^{-2}$ mol.L$^{-1}$, $l$ = 0,1 mm). C) Microscopy images of the aggregates (660 µm x 660 µm) and schematic representation of the proposed aggregation mechanism D) Microscopy images of the gel (660 µm x 660 µm) and schematic representation of the proposed gelation mechanism.

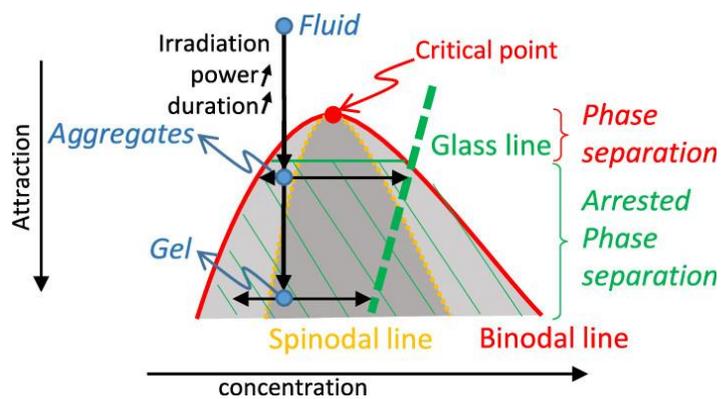

**Scheme 3.** Schematic state diagram illustrating the physical pathway leading to aggregation and gelation. Reprinted with permission from *J. Phys. Chem. B*, **125**, 12063–12071 (2021). Copyright 2021 American Chemical Society.